\documentclass[twocolumn,showpacs,pre,amsmath,amssymb]{revtex4}

\usepackage{graphicx}% Include figure files
\usepackage{dcolumn} % Align table columns on decimal point
\usepackage{bm}      % bold math
\usepackage{latexsym}
\usepackage{color}
\newcommand{\vomega}{\mbox{\boldmath $ \omega $}}
\begin{document}
\title{Inverse cascades and $\alpha$-effect at low magnetic Prandtl number}

\author{P.D. Mininni}
\affiliation{NCAR, P.O. Box 3000, Boulder CO 80307-3000, USA}
\date{\today}

\begin{abstract}
Dynamo action in a fully helical Beltrami (ABC) flow is studied using both 
direct numerical simulations and subgrid modeling. Sufficient scale 
separation is given in order to allow for large-scale magnetic energy 
build-up. Growth of magnetic energy obtains down to a magnetic Prandtl 
number $P_M=R_M/R_V$ close to $0.005$, where $R_V$ and $R_M$ are the 
kinetic and magnetic Reynolds numbers. The critical magnetic Reynolds 
number for dynamo action $R_M^c$ seems to saturate at values close to 
$20$. Detailed studies of the dependence of the amplitude of the 
saturated magnetic energy with $P_M$ are presented. In order to decrease 
$P_M$, numerical experiments are conducted with either $R_V$ or $R_M$ kept 
constant. In the former case, the ratio of magnetic to kinetic energy 
saturates to a value slightly below unity as $P_M$ decreases. Examination 
of energy spectra and structures in real space both reveal that quenching 
of the velocity by the large-scale magnetic field takes place, with an 
inverse cascade of magnetic helicity and a force-free field at large scale 
in the saturated regime.
\end{abstract}

\maketitle
\section{\label{sec:intro}INTRODUCTION}

In recent years the increase in computing power, as well as the 
development of subgrid models for magnetohydrodynamic (MHD) turbulence 
\cite{Muller02a,Muller02b,Ponty04,Mininni05a,Mininni05b} has allowed the 
study of a numerically almost unexplored territory: the regime of low 
magnetic Prandtl number ($P_M=R_M/R_V$, where $R_V$ and $R_M$ are 
respectively the kinetic and magnetic Reynolds numbers). This MHD 
regime is of particular importance since several astrophysical 
\cite{Parker} and geophysical \cite{Roberts01,Kono02} problems are 
characterized by $P_M<1$, as for example in the liquid core of planets 
such as Earth, or in the convection zone of solar-type stars. Also, 
liquid metals (e.g., mercury, sodium, or gallium) used in the laboratory 
in attempting to generate dynamo magnetic fields are in this regime 
\cite{Noguchi02,Petrelis03,Sisan03,Spence05}. 

In recent publications 
\cite{Schekochihin04a,Ponty05,Schekochihin05,Mininni05c,Mininni05d},
driven turbulent MHD dynamos were studied numerically, within the framework 
of rectangular periodic boundary conditions. As $P_M$ is lowered at 
fixed viscosity the magnetofluid becomes more resistive than it is viscous, 
and it was found that magnetic fields were harder to excite by the dynamo 
process because of the increased turbulence in the fluid. The principal 
result was in obtaining the dependence of the critical magnetic Reynolds 
number $R_M^c$ with the magnetic Prandtl number. These studies were done 
for several settings, ranging from coherent helical \cite{Mininni05d} and 
non-helical \cite{Ponty05,Mininni05c} forcing, as well as for random 
forcing \cite{Schekochihin04a,Schekochihin05}. While for coherent 
forcing an asymptotic regime was found at small values of $P_M$, the 
behavior of $R_M^c$ for random forcing is still unclear (see also 
\cite{Vincenzi02,Boldyrev04} for theoretical arguments based on the 
Kazantsev model \cite{Kazantsev68}).

For coherent forcing such as the Taylor-Green vortex (that corresponds to 
several laboratory experiments using two counter-rotating disks), the value 
of $R_M^c$ was observed to increase by a factor larger than six before the 
asymptotic regime for small values of $P_M$ was reached \cite{Ponty05}. 
Although the precise value of $R_M^c$ in experiments is expected to be 
modified by the presence of boundaries, it is of interest to study what 
properties of the forcing can modify and decrease its value. It is well 
known from theory \cite{Moffatt}, two-point closure models \cite{Pouquet76}, 
and direct numerical simulations (DNS) at $P_M=1$ 
\cite{Meneguzzi81,Gilbert88,Brandenburg01} that the presence of net 
helicity in the flow helps the dynamo and decreases the value of $R_M^c$.

In \cite{Mininni05d} dynamos with a helical forcing function were 
studied using the Roberts flow, but mechanical energy was injected at 
a wavenumber $|{\bf k}| = \sqrt{2}$, which left little room in the 
spectrum for any back-transfer of magnetic helicity as expected in the 
helical case \cite{Steenbeck66,Pouquet76,Krause}; indeed, $|{\bf k}|=1$ 
is the only possibility since the computations are done in a box of length 
$2\pi$ corresponding to a $k=1$ gravest mode. In this work, in contrast, 
we study the effect of a fully helical Arn'old-Childress-Beltrami (ABC) 
forcing \cite{Childress} with energy injected at a slightly smaller 
scale (note that the ABC forcing is related to the Roberts forcing, 
since it can be defined as a superposition of three Roberts flows). 
As a result of the intermediate scale forcing, some $\alpha$-effect 
or inverse cascade of magnetic helicity can a priori develop and a 
magnetic field at large scales can grow. 

ABC flows and helical dynamos were explored in many different contexts 
in the literature (see e.g. \cite{Galanti92} for a study close to $P_M=1$, 
and \cite{Ponty95,Hollerbach95} for studies in the context of fast dynamo 
action). The main aim of the present work is to study the impact of 
helical flows at intermediate scales in the development of magnetic 
fields through dynamo action at $P_M<1$. In this context, it is worth 
noting that some simulations of ABC dynamos in the low magnetic Prandtl 
number regime were discussed in Refs. \cite{Brandenburg01,Archontis03}, 
although no systematic exploration of the space of parameters was 
attempted. Also, Ref. \cite{Mininni05e} presented some preliminary 
results for the kinematic dynamo regime with ABC forcing. In this 
work we will focus on the study of the generation of large scale 
magnetic fields and of the non-linear saturation regime. A similar 
study was recently conducted in Ref. \cite{Frick06} using mean field 
theory \cite{Steenbeck66,Krause} and shell models. To the best of our 
knowledge, ours is the first attempt to systematically study the 
saturation values of the fields for helical flows at $P_M<1$ in 
numerical simulations.

\section{\label{sec:equations}DEFINITIONS AND METHODOLOGY}

In a familiar set of dimensionless (``Alfv\'enic'') units the 
equations of magnetohydrodynamics are:
\begin{eqnarray}
\frac{\partial {\bf v}}{\partial t} + {\bf v \cdot \nabla v} &=& 
    -{\bf \nabla} {\mathcal P} + {\bf j \times B} + \nu \nabla^2 {\bf v} + 
    {\bf f} , \label{eq:momentum} \\
\frac{\partial {\bf B}}{\partial t} + {\bf v \cdot \nabla B} &=&
    {\bf B \cdot \nabla v} + \eta \nabla^2 {\bf B} ,
    \label{eq:induc}
\end{eqnarray}
with ${\bf \nabla \cdot v} = {\bf \nabla \cdot B} = 0$. Here, ${\bf v}$ 
is the velocity field, regarded as incompressible, and ${\bf B}$ is the 
magnetic field, related to the electric current density ${\bf j}$ by 
${\bf j}={\bf \nabla \times B}$. ${\mathcal P}$ is the pressure, 
obtained by solving the Poisson equation that results from taking 
the divergence of Eq. (\ref{eq:momentum}) and using the 
incompressibility condition ${\bf \nabla \cdot v} = 0$. The viscosity 
$\nu$ and magnetic diffusivity $\eta$ define mechanical Reynolds numbers 
and magnetic Reynolds numbers respectively as $R_V = LU/\nu$ and 
$R_M = LU/\eta$. Here $U$ is a typical turbulent flow speed (the 
r.m.s. velocity in the following sections, $U = \left<u^2\right>^{1/2}$, 
with the brackets denoting spatial averaging), and $L$ is a length scale 
associated with spatial variations of the large-scale flow (the integral 
length scale of the flow). We can also define a Taylor based Reynolds 
number $R_\lambda = \lambda U/\nu$, where $\lambda$ is the Taylor 
lengthscale, defined below.

Some global quantities will appear repeatedly in the next sections. These 
are the total energy (the sum of the kinetic $E_V$ and magnetic $E_M$ 
energies) $E = E_V + E_M ={{1}\over{2}} \int ({\bf u}^2 + {\bf B}^2) \, dV$, 
the magnetic helicity $H_M = \int {\bf A} \cdot {\bf B} \, dV$ (where 
${\bf A}$ is the vector potential, defined such as 
${\bf B} = \nabla \times {\bf A}$), and the kinetic helicity 
$H_V = \int {\bf v} \cdot \vomega \, dV$ (where 
$\vomega = \nabla \times {\bf v}$ is the vorticity). While $E$ and 
$H_M$ are ideal ($\nu = \eta = 0$) quadratic invariants of the MHD 
equations, $H_K$ is not. In practice, kinetic helicity in helical dynamos 
is injected into the flow by the mechanical forcing ${\bf f}$ (e.g., by 
rotation and stratification  in geophysical and astrophysical flows 
\cite{Moffatt}).

Equations (\ref{eq:momentum}) and (\ref{eq:induc}) are solved 
numerically using a parallel pseudospectral code, as described 
in Refs. \cite{Mininni05c,Mininni05d}. We impose rectangular periodic 
boundary conditions throughout, using a three-dimensional box of 
edge $2\pi$. The integral and Taylor scales are defined respectively 
as
\begin{equation}
L = 2 \pi \sum_{\bf k}{k^{-1} |\hat{\bf v}({\bf k})|^2} \bigg/ 
    \sum_{\bf k}{|\hat{\bf v}({\bf k})|^2} ,
\label{eq:integral}
\end{equation}
\begin{equation}
\lambda = 2 \pi \left( \sum_{\bf k}{|\hat{\bf v}({\bf k})|^2} \bigg/ 
    \sum_{\bf k}{k^2 |\hat{\bf v}({\bf k})|^2} \right)^{1/2} ,
\label{eq:taylor}
\end{equation}
where $\hat{\bf v}({\bf k})$ is the amplitude of the mode with wave 
vector ${\bf k}$ ($k=|{\bf k}|$) in the Fourier transform of ${\bf v}$.

The external forcing function ${\bf f}$ in Eq. (\ref{eq:momentum}) 
injects both kinetic energy and kinetic helicity. For ${\bf f}$ 
we use the ABC flow
{\setlength\arraycolsep{2pt}
\begin{eqnarray}
{\bf f}_{\rm ABC} &=& f_0 \left\{ \left[B \cos(k_F y) + 
    C \sin(k_F z) \right] \hat{x} + \right. {} \nonumber \\
&& {} + \left[A \sin(k_F x) + C \cos(k_F z) \right] \hat{y} + 
   {} \nonumber \\
&& {} + \left. \left[A \cos(k_F x) + B \sin(k_F y) \right] 
   \hat{z} \right\},
\label{eq:ABC}
\end{eqnarray}}
with $A=0.9$, $B=1$, $C=1.1$ \cite{Archontis03}, and $k_F = 3$. The 
ABC flow is an eigenfunction of the curl with eigenvalue $k_F$, and 
as a result if used as an initial condition it is an exact solution 
of the Euler equations. In the hydrodynamic simulations, for large 
enough $\nu$ (small $R_V$) the laminar solution is stable. As $\nu$ 
is decreased the laminar flow becomes unstable and develops turbulence 
(see \cite{Podvigina94} for a study of the early bifurcations at 
intermediate Reynolds numbers).

To properly resolve the turbulent flow, the maximum wavenumber in the 
code $k_{max}=N/3$ ($N$ is the linear resolution and the standard 
$2/3$-rule for dealiasing is used) has to be smaller than the mechanic 
dissipation wavenumber $k_\nu = (\epsilon/\nu^3)^{1/4}$ 
($\epsilon \sim U^3/L$ is the energy injection rate). As a result, as 
$\nu$ decreases and $R_V$ increases, the linear resolution $N$ has to 
be increased. At some point the use of DNS to solve Eqs. 
(\ref{eq:momentum}) and (\ref{eq:induc}) turns to be too expensive 
from the computational point of view and some kind of model for 
unresolved scales is needed. 

To extend the range of $R_V$ and $P_M$ studied, we use the Lagrangian 
average MHD equations (LAMHD, also known as the MHD $\alpha$-model) 
\cite{Holm02a,Holm02b,Mininni05a}
\begin{eqnarray}
\frac{\partial {\bf v}}{\partial t} + {\bf u}_s \cdot \nabla {\bf v} &=& 
    - v_j \nabla u_s^j -\nabla \widetilde{{\cal P}} + {\bf j} \times 
    {\bf B}_s \nonumber \\
    {} && + \nu \nabla^2 {\bf v} + {\bf f}, \label{eq:alpNS} \\
\frac{\partial {\bf B}_s}{\partial t} + {\bf u}_s \cdot \nabla {\bf B}_s 
    &=& {\bf B}_s \cdot \nabla {\bf u}_s + \eta \nabla^2 {\bf B} . 
    \label{eq:alpind} 
\end{eqnarray}
In these equations, the pressure $\widetilde{{\cal P}}$ is determined, 
as before, from the divergence of Eq. (\ref{eq:alpNS}) and the 
incompressibility condition. The subindex $s$ denotes smoothed 
fields, related to the unsmoothed fields by
\begin{eqnarray}
{\bf v} &=& \left(1 - \alpha_V^2 \nabla^2\right) {\bf u}_s \\
{\bf B} &=& \left(1 - \alpha_B^2 \nabla^2\right) {\bf B}_s . 
\end{eqnarray}
The total energy in this system is given by 
$E=E_V+E_M={{1}\over{2}}\int ({\bf v}\cdot {\bf u}_s +{\bf B}\cdot {\bf B}_s)\, dV$; 
it is one of the ideal quadratic invariants of the LAMHD equations. 
Equivalently, the magnetic helicity invariant is now given by 
$H_M = \int {\bf A}_s \cdot {\bf B}_s dV$, where the smooth vector 
potential is defined such as ${\bf B}_s = \nabla \times {\bf A}_s$. 
The expression for the kinetic helicity is the same in MHD and LAMHD.

The LAMHD equations are a regularization of the MHD equations, and as 
a result they allow for simulations of turbulent flows at a given Reynolds number using a lower resolution 
than in DNS. This subgrid model was validated against DNS of MHD flows 
in \cite{Mininni05a,Mininni05b}. As in previous studies of dynamo action 
at low $P_M$, the ratio of the two filtering scales $\alpha_V$ and 
$\alpha_B$ was set using the ratio of the kinetic and magnetic 
dissipation scales, i.e. $\alpha_V/\alpha_B = P_M^{3/4}$ \cite{Ponty05}. 
The value of $\alpha_V$ depends on the linear resolution and was 
adjusted to $1/\alpha_V \approx k_{max}/2$ \cite{Geurts06}.

\begin{table}
\caption{\label{table:runs}Parameters for the simulations: kinematic 
viscosity $\nu$, Taylor Reynolds number $R_\lambda$, mechanical Reynolds 
number $R_V$, range of values of the magnetic Reynolds number $R_M$ for 
a given flow, linear resolution $N$, and value of the mechanic filter 
length $\alpha_V$ (with $\alpha_V/\alpha_B = [R_M/R_V]^{3/4}$ in each 
case). For direct simulations (no subgrid model), $\alpha_V$ need not be 
defined. Runs in set 6a have the same values of parameters than in set 6, 
but a subgrid model at a lower resolution was used. Run 9 has the same 
parameters as set 9b with $R_M=41$ but was done using a DNS. The lowest 
$P_M$ achieved for this set of runs is $\approx 0.005$.}
\begin{ruledtabular}
\begin{tabular}{lcccccc}
Set & $\nu$          & $R_\lambda$ & $R_V$ & $R_M$  & $N$ & $\alpha_V$ \\
\hline
1   &$0.2$           & 11          & 11    & 9--16  & 64  & -        \\
2   &$0.1$           & 21          & 23    & 10--19 & 64  & -        \\
3   &$4\times10^{-2}$& 55          & 71    & 17--71 & 64  & -        \\
4   &$9\times10^{-3}$& 161         & 240   & 18--54 & 64  & -        \\
5   &$4\times10^{-3}$& 250         & 450   & 15--450& 128 & -        \\
6   &$2\times10^{-3}$& 360         & 820   & 10--41 & 256 & -        \\
6a  &$2\times10^{-3}$& 290         & 840   & 10--41 & 64  & $0.1$    \\
7   &$1\times10^{-3}$& 340         & 1700  & 14--42 & 128 & $0.0625$ \\
8   &$6.2\times10^{-4}$& 680       & 2500  & 39     & 512 & -        \\
9   &$5\times10^{-4}$& 500         & 3400  & 41     & 512 & -        \\
9b  &$5\times10^{-4}$& 500         & 3400  & 14--42 & 256 & $0.03125$\\
10&$2.5\times10^{-4}$& 1100        & 6200  & 77     & 1024& -        \\
\end{tabular}
\end{ruledtabular}
\end{table}

In the next section, we describe the computations and the results for 
both the kinematic dynamo regime [where ${\bf j \times B}$ is 
negligible in Eq. (\ref{eq:momentum})], and for full MHD (where the 
Lorentz force modifies the flow). The first step is to establish what 
are the thresholds in $R_M$ at which dynamo behavior sets in as $R_V$ 
is raised and $P_M$ is decreased (Section \ref{sec:RMC}). The procedure 
to do this is the following (see e.g., Ref. \cite{Ponty05}). First a 
hydrodynamic simulation at a given value of $R_V$ is done. Then, a 
small and random (non-helical) magnetic field is introduced, and 
several simulations are done changing only the value of $R_M$. At 
a given $R_M$, the magnetic energy can either decay or grow 
exponentially. In each simulation, the growth rate $\sigma$ is then 
defined as $\sigma = d \ln(E_M) / dt$. The critical magnetic Reynolds 
number $R_M^c$ for the onset of dynamo action corresponds to $\sigma=0$, 
and in practice is obtained from a linear interpolation between the two 
points with respectively positive and negative $\sigma$ closest to zero. 
The growth rate $\sigma$ is typically expressed in units of the 
reciprocal of the large-scale eddy turnover time $T=L/U$.

\begin{figure}
\includegraphics[width=9cm]{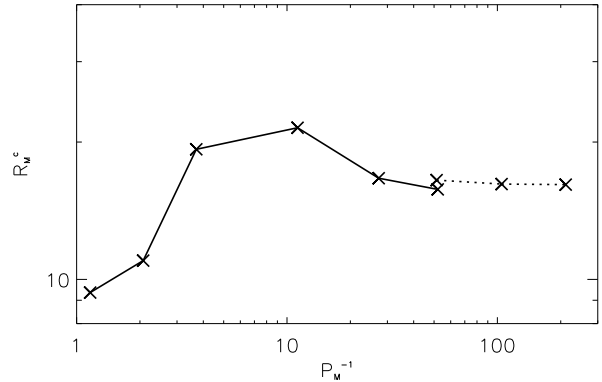}
\caption{Critical magnetic Reynolds $R_M^c$ as a function of $P_M^{-1}$: 
     DNS (solid line) and LAMHD (dotted line). Note the saturation for 
     $P_M \le 0.02$.}
\label{fig:crmm}
\end{figure}

Once the values of $R_M^c$ for different values of $P_M \le 1$ have 
been found, simulations for $R_M > R_M^c$ are conducted for longer 
times (Section \ref{sec:time}). In this case, magnetic fields are 
initially amplified exponentially, and then saturate due to the back 
reaction of the magnetic field on the flow. In helical flows, this 
saturation is accompanied by the growth of magnetic fields in the 
largest scale available in the box. In this regime, we will study 
the maximum value attained by the magnetic energy as a function of 
$P_M$ (Section \ref{sec:satura}), as well as the amount of magnetic 
energy at scales larger than the forcing scale (Section 
\ref{sec:spectrum}). Finally, Section 4 is the conclusion.

\begin{figure}
\includegraphics[width=9cm]{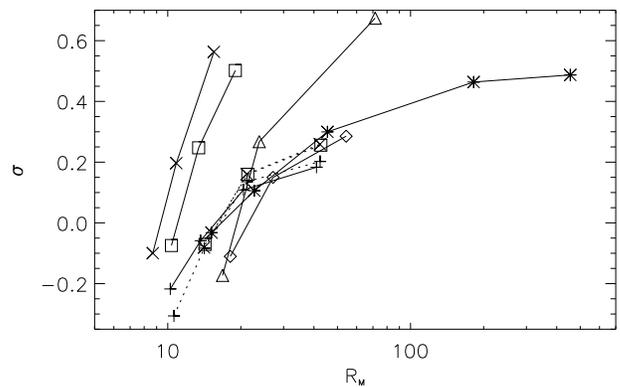}
\caption{Growth rates as a function of $R_M$. Each line corresponds 
    to several simulations at constant $R_V$ (fixed $\nu$), and each 
    point in the line indicates the exponential growth (or decay) rate 
    at a fixed value of $R_M$. The point where each curve crosses 
    $\sigma=0$ gives the threshold $R_M^c$ for dynamo instability. 
    Points from DNS are connected with solid lines, and labels are: set 
    1 ($\times$), set 2 ($\square$), set 3 ($\bigtriangleup$), set 4 
    ($\diamond$), set 5 ($*$), and set 6 ($+$). Points from LAMHD 
    simulations are connected with dotted lines: set 6a ($+$), set 7  
    ($\times$), and set 9b ($\square$). Note the accumulation of lines 
    near $R_M\approx 20$.}
\label{fig:growth}
\end{figure}

\section{\label{sec:results}SIMULATIONS AND RESULTS}

In order to obtain a systematic study of dynamo action for ABC forcing 
and $P_M \le 1$, a suite of  several simulations was conducted. 
Table \ref{table:runs} shows the parameters used in the simulations. 
Note that when a range is invoked in the values of $R_M$, it indicates 
several runs were done with the same value of $R_V$ but changing the 
value of $R_M$ to span the range (typically three to five runs). The 
set of runs 6 and 6a have the same parameters ($\nu$, $\eta$, and r.m.s 
velocity), but while set 6 comprises DNS at resolutions of $256^3$ 
grid points, in set 6a the spatial resolution is $64^3$ and the LAMHD 
equations were used in order to further the testing of the model. Similar 
considerations apply to run 9 and set 9b.

\subsection{\label{sec:RMC}Threshold for dynamo action}

\begin{figure}
\includegraphics[width=9cm]{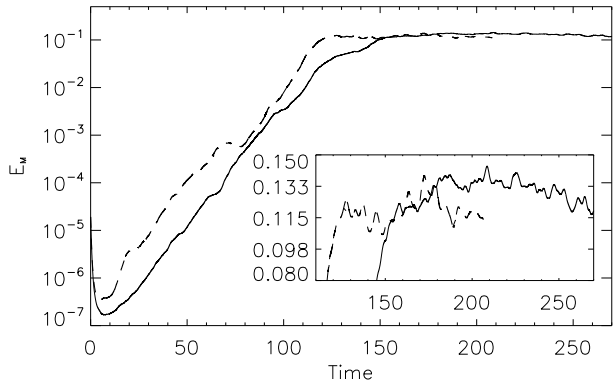}
\caption{Time history of the magnetic energy for runs in set 6 (dashed 
    line) and in set 6a (solid line) with $R_M\approx 41$ and 
    $P_M=5 \times 10^{-2}$. The inset shows the time evolution of the 
    magnetic energy after the nonlinear saturation, in linear scale.}
\label{fig:energ}
\end{figure}

Figure \ref{fig:crmm} summarizes the results of the study of the 
dependence of the threshold $R_M^c$ as $P_M$ is decreased. For values 
of $R_M$ above the curve, dynamo action takes place and initially small 
magnetic fields are amplified. Below the curve, Ohmic dissipation is too 
large to sustain a dynamo. Noteworthy is the qualitative similarity of 
the curve between the ABC flow and previous results using different 
mechanical forcings \cite{Ponty05,Mininni05c,Mininni05d}. Namely, an 
increase in $R_M^c$ is observed as turbulence develops, and then an 
asymptotic regime is found in which the value of $R_M^c$ is independent 
of $P_M$. Note that LAMHD simulations were used to extend the study for 
values of $P_M$ smaller than what can be studied using DNS. Simulations 
at the same value of $P_M$ were carried with the two methods to validate 
the results from the subgrid model (sets 6 and 6a). This procedure was 
used before in Ref. \cite{Ponty05}. As in the previous study, the LAMHD 
equations slightly overestimate the value of $R_M^c$.

\begin{figure}
\includegraphics[width=9cm]{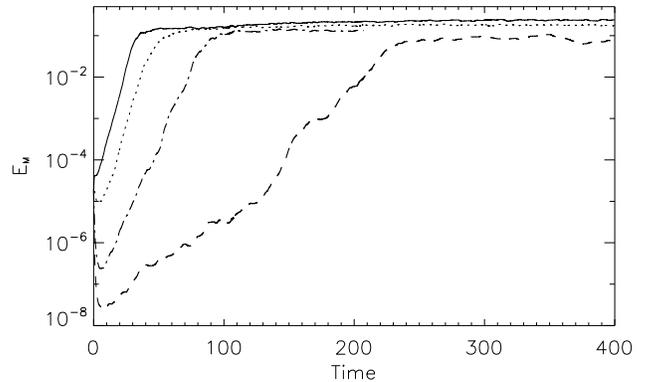}
\caption{Time history of the magnetic energy for runs in set 5 
    (constant $R_V \approx 450$). The magnetic Reynolds number in 
    each run is $R_M \approx 22$ (dashed line), $R_M \approx 45$ 
    (dash-dotted line), $R_M \approx 180$ (dotted line), and 
    $R_M \approx 450$ (solid line).}
\label{fig:RVene}
\end{figure}

Besides the similarities in the shape of the curves for different 
forcing functions, two quantitative differences are striking: (i) only 
a mild rise in $R_M^c$ is observed here as $P_M$ is decreased (a factor 
2, while a factor larger than 6 obtains for the Taylor-Green vortex 
\cite{Ponty05}); and (ii) the asymptotic value of $R_M^c$ for small 
values of $P_M$ is ten times smaller than for other flows studied 
\cite{Ponty05,Mininni05d}. A similar result was obtained using mean 
field theory and shell models in Ref. \cite{Frick06}, and the 
quantitative differences observed were associated with the relative 
ease to excite large-scale helical dynamos compared with non-helical 
and small-scale dynamos.

Note that the curve in Fig. \ref{fig:crmm} was constructed using the 
sets 1 to 7 and 9b of Table 1. Several runs at constant $R_V$ but 
varying $R_M$ are required to define $R_M^c$. Set 9b reveals a dynamo 
at the lowest magnetic Prandtl number known today in numerical 
simulations, namely $P_M=4.7 \times 10^{-3}$.

\begin{figure}
\includegraphics[width=9cm]{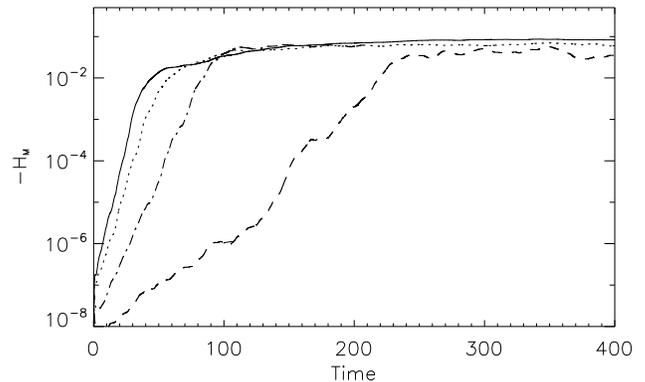}
\caption{Time history of (minus) the magnetic helicity for runs in 
    set 5 (constant $R_V \approx 450$). The magnetic Reynolds number 
    in each run is $R_M \approx 22$ (dashed line), $R_M \approx 45$ 
    (dash-dotted line), $R_M \approx 180$ (dotted line), and 
    $R_M \approx 450$ (solid line).}
\label{fig:RVhel}
\end{figure}

Figure \ref{fig:growth} shows the details of how the thresholds for 
the determination of the $R_M^c=f(P_M^{-1})$ curve were calculated. 
For small initial $E_M$, broadly distributed over a set of wavenumbers, 
$\eta$ was decreased in steps to raise $R_M$ in the same mechanical 
setting until a value of $\sigma \approx 0$ was identified. A linear 
fit between the two points with $\sigma$ closest to $0^{\pm}$ provides 
a single point on the curves in Fig. \ref{fig:crmm}. Note that Figure 
\ref{fig:growth} also gives bounds for the uncertainties in the 
determination of the threshold $R_M^c$ (see e.g. Ref. \cite{Ponty05}): 
errors in Fig. \ref{fig:crmm} can be defined as the distance between 
the value of $R_M^c$ and the value of $R_M$ in the simulation with 
$\sigma$ closest to 0. Note also the asymptotic approach to a growth 
rate of order unity for large values of the magnetic Reynolds number, 
as for example in the runs in set 5.

\subsection{\label{sec:time}Time evolution}

A comparison of the time evolution of the magnetic energy in two 
dynamo runs with the same mechanic and magnetic Prandtl number 
($R_M \approx 41$, $P_M = 5\times 10^{-2}$) is shown in Fig. 
\ref{fig:energ}. One of the runs is a DNS from set 6, while the other 
is a LAMHD simulation from set 6a. Two different stages can be 
identified at first sight in these runs. The kinematic regime at 
early times, with an exponential amplification of the magnetic energy 
(used to define the growth rates and thresholds in Figs. \ref{fig:crmm} 
and \ref{fig:growth}), and the nonlinear saturated regime at late times. 
As expected from the results discussed in the previous subsection, the 
LAMHD equations at a coarser grid ($64^3$) are able to capture the 
kinematic dynamo regime. While in the DNS with a resolution of $256^3$ 
the growth rate is $\sigma \approx 0.18$, in the LAMHD simulation 
$\sigma \approx 0.20$. But the LAMHD simulation also captures 
properly the nonlinear saturation (albeit the saturated level 
is reached a bit earlier) and the amplitudes of the magnetic energy in 
the steady state are comparable (see insert in Fig. \ref{fig:energ}). 
Small differences observed in the time evolution are likely due to 
differences in the initial random magnetic seed. In the following, 
we shall use both DNS and LAMHD simulations to study the nonlinear 
saturated regime at low $P_M$.

\begin{figure}
\includegraphics[width=9cm]{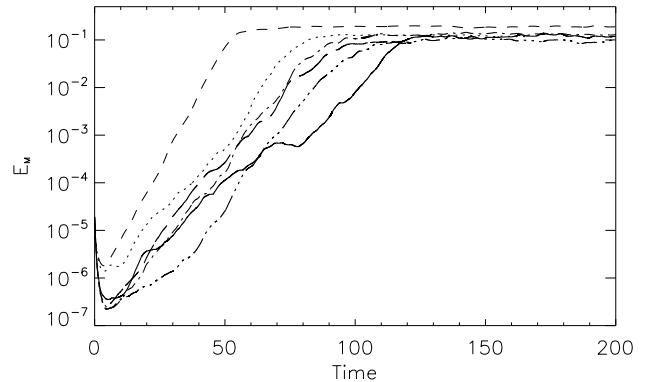}
\caption{Time history of the magnetic energy in simulations at 
    constant $R_M \approx 41$ ($\eta = 4 \times 10^{-2}$). The 
    different runs are taken from set 3 ($P_M = 1$, dash), set 4 
    ($P_M = 0.225$, dot), set 5 ($P_M = 0.1$, dash-dot), 
    set 6 ($P_M = 0.05$, solid), set 7 ($P_M = 0.025$, dash-triple 
    dot), and finally set 9 ($P_M = 0.0125$, long dash).}
\label{fig:RMene}
\end{figure}

In helical flows, as magnetic energy saturates, a large scale magnetic 
field develops (i.e., at scales larger than the forcing scale) due to 
the helical $\alpha$-effect 
\cite{Steenbeck66,Pouquet76,Moffatt,Krause,Brandenburg01,Brandenburg05}. 
It is of interest to know what happens with the amplitude of the magnetic 
field as the value of $P_M$ is decreased. An example is shown in Figure 
\ref{fig:RVene}, which gives the magnetic energy as a function of time 
for runs in set 5. Only the value of $R_M$ (and therefore of $P_M$) is 
changed between the runs ($R_V \approx 450$ in all runs and $P_M$ varying 
from 1 to $0.03$). For large values of $R_M$ (but not necessarily for 
values of $P_M$ close to unity), the growth rate $\sigma$ is independent 
of $R_M$ and of order one as noted in Section \ref{sec:RMC}. Furthermore, 
as $P_M$ is decreased, both $\sigma$ and the saturation value of the 
magnetic energy decrease. However, for the lowest value of $P_M$, the 
magnetic Reynolds number is quite low and in that context computations 
at a higher value of $R_V$ and with the same sets of $P_M$ are of value 
to see what fraction of the present result is a threshold effect at low 
$R_M$.

\begin{figure}
\includegraphics[width=9cm]{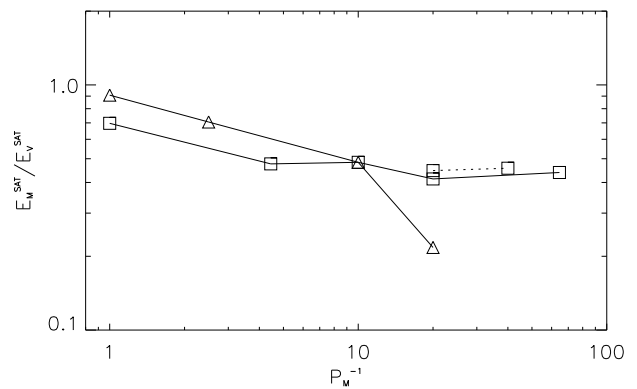}
\caption{Saturation value of the magnetic energy (normalized by the 
     kinetic energy in the saturated regime). The triangles correspond 
     to simulations at constant $R_V$, while the squares correspond to 
     simulations at constant $R_M$ (squares connected with solid lines 
     are from DNS, while squares connected with dotted lines are from 
     LAMHD simulations). Note the saturation at low $P_M$ for constant 
     $R_M$ runs.}
\label{fig:scene}
\end{figure}

Figure \ref{fig:RVhel} shows the evolution of the magnetic 
helicity as a function of time for the same simulations than in Fig. 
\ref{fig:RVene}. The external forcing injects positive kinetic 
helicity in the flow. In the kinematic regime, the $\alpha$ effect 
is proportional to minus the kinetic helicity \cite{Steenbeck66,Krause}. 
From mean field theory, the magnetic field in the large scales should 
grow with magnetic helicity of the same sign than the $\alpha$ effect 
(negative), as indeed observed (see Refs. \cite{Brandenburg01,Mininni03} 
for helical dynamo simulations at $P_M = 1$). In the simulations, 
magnetic helicity grows exponentially during the kinematic regime. 
In runs with small $R_M$, the saturated state is reached shortly 
after the saturation of the exponential phase. But as $R_M$ is 
increased, it is now clear that an intermediate stage develops in which 
magnetic energy and helicity keep growing slowly. As a result, 
saturation takes place in longer times, and the time to reach the 
final steady state depends on the large scale magnetic diffusion 
time ($T_\eta \approx 4 \pi^2/\eta$). The dependence of the 
saturation time with $R_M$ can be observed in Fig. \ref{fig:RVhel}. 
It is also worth mentioning that even in the runs with $P_M<1$, the 
saturation of magnetic helicity can be well described by the formula 
$H_M(t) \sim 1-\exp[-2\eta k_0^2 (t-t_{\rm sat})]$, where $k_0=1$ 
is the gravest mode, and $t_{\rm sat}$ is the saturation time of 
the small scale magnetic field \cite{Brandenburg01}. This indicates that
the slow saturation of the dynamo is dominated by the evolution of 
the magnetic helicity in the largest scale in the system.

\begin{figure}
\includegraphics[width=9cm]{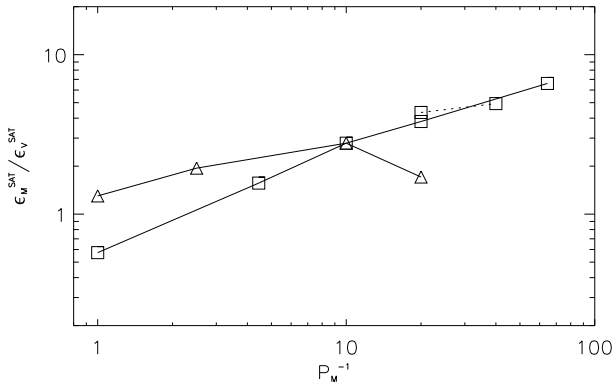}
\caption{Saturation value of the magnetic dissipation rate normalized 
     by the kinetic energy in the saturated regime. The triangles 
     correspond to simulations at constant $R_V$, while the squares 
     correspond to simulations at constant $R_M$ (squares connected 
     with solid lines are from DNS, while squares connected with dotted 
     lines are from LAMHD simulations).}
\label{fig:disrate}
\end{figure}

From Figs. \ref{fig:RVene} and \ref{fig:RVhel} it seems apparent that 
small values of $P_M$ have a negative impact on the amplitude of the 
magnetic field generated by the dynamo. However, different results 
are obtained when the space of parameters is explored keeping $R_M$ 
constant and increasing $R_V$, as another way to decrease $P_M$. 
Figure \ref{fig:RMene} shows the results in this case for the time 
evolution of the magnetic energy. As $R_V$ is increased from small 
values, a drop in the growth rate $\sigma$ and in the saturation value 
of the magnetic energy is observed. But then an asymptotic regime is 
reached, in which both $\sigma$ and the saturation value seem to be 
roughly independent of $R_V$ and $P_M$. As a result, we conclude that 
the behavior observed in Figs. \ref{fig:RVene} and \ref{fig:RVhel} is 
the result of critical slowing down: if the space of parameters is 
explored at constant $R_V$, as $P_M$ is decreased $R_M$ gets closer 
to $R_M^c$ until no dynamo action is possible. On the other hand, 
all the simulations with $P_M \le 0.05$ shown in Fig. \ref{fig:RVhel} 
have $R_M/R_M^c$ approximately constant (see Fig. \ref{fig:crmm}) and 
critical slowing down is not observed. However, we note that the value 
of $R_M$ for these runs is still modest.

\subsection{\label{sec:satura}Saturation values}

\begin{figure}
\includegraphics[width=9cm]{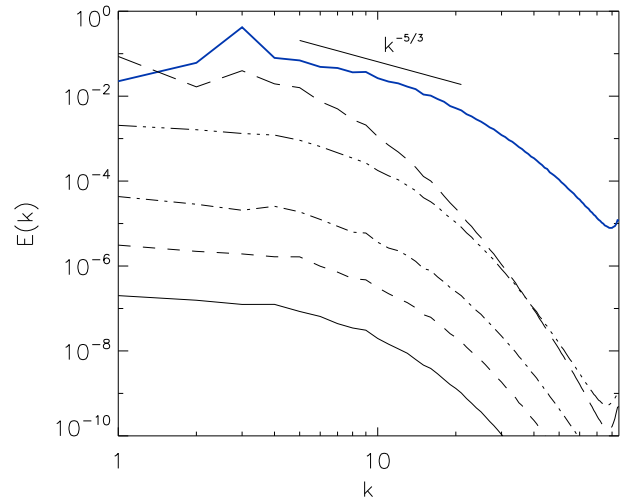}
\caption{(Color online) Kinetic energy spectrum at $t=0$ [thick (blue) 
    lines], and magnetic energy spectrum (thin lines) at different 
    times: $t=11$ (solid line), $t=29$ (dashed line), $t=47$ (dash-dotted 
    line), $t=95$ (dash-triple dotted line), $t=120$ (long dashed line). 
    The spectra are for a run in set 6 with $R_M \approx 41$. The last 
    time is in the saturation regime (see Fig. \ref{fig:RMene}).}
\label{fig:spectrum}
\end{figure}

The amplitude of the magnetic energy (normalized by the kinetic energy), 
after the nonlinear saturation takes place, as a function of the magnetic 
Prandtl number is shown in Fig. \ref{fig:scene}. This figure summarizes 
the results discussed in Figs. \ref{fig:RVene} and \ref{fig:RMene}. As 
the value of $P_M$ is decreased, if $R_V$ is kept constant and $R_M$ 
(and thus $P_M$) decreases, the saturation of the dynamo takes place 
for lower values of the magnetic energy. This is to be expected since 
as we decrease $P_M$ we also decrease $R_M$ and at some point $R_M^c$ 
is reached. It is not clear whether such a strong dependence would be 
observed if the constant $R_V$ runs were performed at substantially 
higher values of $R_V$ as found in astronomical bodies and in the 
laboratory; however, such runs would be quite demanding from a 
numerical standpoint unless one resorts to LES (Large Eddy Simulations) 
techniques, few of which have been developed in MHD (see e.g., 
\cite{Muller02a,Muller02b,Ponty04}). For values of $R_M$ smaller than 
$R_M^c$, no dynamo action is expected and the ratio $E_M/E_V$ should 
indeed go to zero. On the other hand, in the simulations with constant 
$R_V$, the ratio $E_M/E_V$ seems to saturate for $P \lesssim 0.25$ and 
reach an approximately constant value close to $\approx 0.5$. This 
indicates that small scale turbulent fluctuations in the velocity field 
are strongly quenched by the large scale magnetic field, as will be also 
shown later in the spectral evolution of the energies. The ratio 
$E_M/E_V$ in helical large-scale dynamos is also expected to be 
dependent on the scale separation between the forcing wavenumber (here 
fixed to $k_F = 3$) and the largest wavenumber in the system (here $k=1$). 
As the scale separation increases and there is more space for an inverse 
cascade of magnetic helicity, we expect the ratio $E_M/E_V$ in the 
$P_M<1$ regime to also increase.

\begin{figure}
\includegraphics[width=9cm]{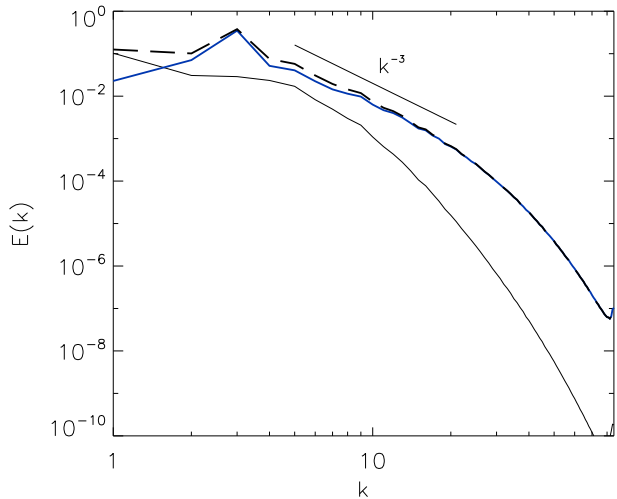}
\caption{(Color online) Kinetic [thick (blue) line] and magnetic 
    energy spectra (thin line) at $t=210$ in the simulation in set 
    6 with $R_M \approx 41$ ($P_M = 0.05$). The thick dashed line 
    shows the total energy spectrum.} 
\label{fig:satura1}
\end{figure}

Figure \ref{fig:disrate} shows the ratio of the magnetic energy 
dissipation rate $\epsilon_M = \eta \left< {\bf j}^2 \right>$ to 
the kinetic energy dissipation rate 
$\epsilon_V = \nu \left< \vomega^2 \right>$ in the saturated 
state for the same runs than in Fig. \ref{fig:scene} (in the LAMHD 
equations, the dissipation rates are 
$\epsilon_M = \eta \left< {\bf j}^2 \right>$ and 
$\epsilon_V = \nu \left< \vomega \cdot \vomega_s \right>$, where 
$\vomega_s = \nabla \times {\bf u}_s$ \cite{Mininni05b}). At constant 
$R_V$, for small values of $P_M$, critical slow down is again observed, 
as the value of $R_M$ gets closer to the threshold. On the other hand, 
at constant $R_M$, more and more energy is dissipated by Ohmic 
dissipation as $P_M$ is decreased.

\subsection{Spectral evolution \label{sec:spectrum}}

\begin{figure}
\includegraphics[width=9cm]{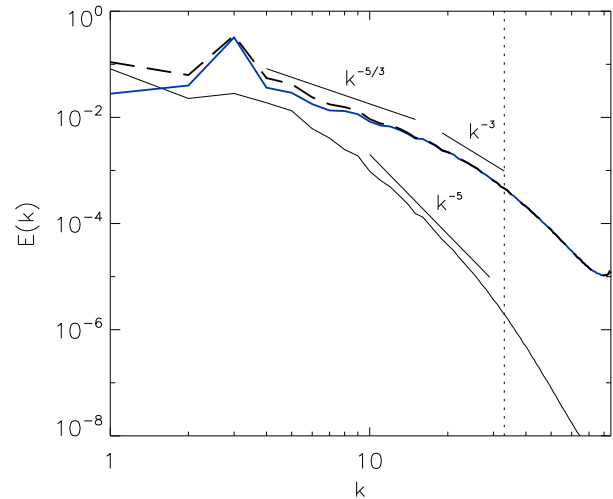}
\caption{(Color online) Kinetic [thick (blue) line] and magnetic 
    energy spectra (thin line) at $t=150$ in the simulation in set 
    9 with $R_M \approx 41$ ($P_M = 0.0125$) in the saturated regime. 
    The thick dashed line shows the total energy spectrum and the thin 
    vertical dotted line the wavenumber at which the $\alpha$ filtering 
    sets in. Note the compatibility of the spectra with a Kolmogorov law 
    in the large scales for the kinetic spectrum, followed by a steeper 
    power law.} 
\label{fig:satura2}
\end{figure}

In Refs. \cite{Mininni05c,Mininni05d} it was shown using different 
forcing functions that even at low $P_M$ the magnetic energy spectrum 
in the kinematic regime of the dynamo peaks at small scales. In these 
simulations, the critical magnetic Reynolds number $R_M^c$ was of the 
order of a few hundreds, and as a result small scales were excited. 
For ABC forcing, $R_M^c$ is of the order of a few tens and close to 
the threshold small scales are damped fast. Only large-scale dynamo 
action is observed and thus, even at early times, the magnetic energy 
spectrum peaks at large scales. However, if $R_M$ is increased above 
$R_M \approx 400$, a magnetic energy spectrum that peaks at scales 
smaller than the forcing scale (as in Refs. \cite{Mininni05c,Mininni05d}) 
is recovered. We focus here on large-scale dynamo action, and as a result 
will discuss the spectral evolution in simulations with $R_M$ of a few 
tens.

Figure \ref{fig:spectrum} shows the evolution of the magnetic energy 
spectrum at different times for a run in set 6 with $R_M \approx 41$. 
As in previous studies, in the kinematic regime all the Fourier shells 
grow with the same rate. Then, magnetic saturation is reached and the 
mode at $k=1$ keeps growing until it eventually saturates itself. Figure 
\ref{fig:satura1} shows the kinetic and magnetic energy spectra at late 
times ($t=210$) after nonlinear saturation in the simulation in set 6 
with $R_M \approx 41$ ($P_M = 0.05$). At $k=1$ the system is dominated by 
magnetic energy, but at smaller scales the magnetic energy spectrum 
drops fast. The kinetic energy spectrum peaks at the forcing band 
($k=3$) and then drops with a slope compatible with $k^{-3}$. This drop 
is due to the action of the Lorentz force that removes mechanical 
energy from the $k=3$ shell to sustain the magnetic field at $k=1$ 
\cite{Alexakis05a,Mininni05f}.

\begin{figure}
\includegraphics[width=9cm]{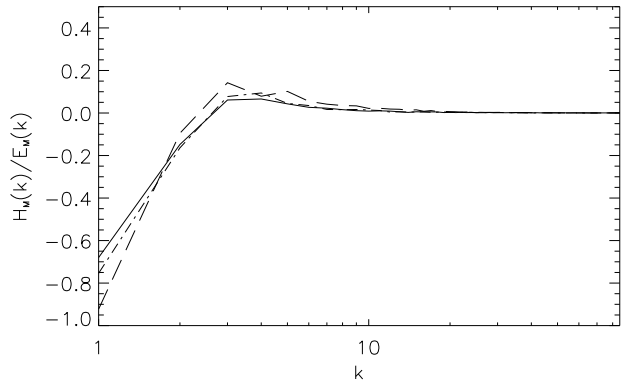}
\caption{Spectrum of relative magnetic helicity $k^{-1} H_M(k)/E_M(k)$ at 
    different times in the simulation in set 6 with $R_M \approx 41$ 
    ($P_M = 0.05$). The labels are as in Fig. \ref{fig:spectrum}. 
    Note the evolution towards a force-free field at $k=1$, the small 
    excess of positive helicity at scales slightly smaller than the forcing 
    scale, and the absence of relative magnetic helicity in the small scales 
    at all times.}
\label{fig:helspec}
\end{figure}

A slope close to a $k^{-3}$ power law in the kinetic energy spectrum 
in the saturated regime at small scales is observed in several of the 
simulations with $P_M<1$. Simulations with small $P_M$ and larger 
values of $R_V$ were done using both the LAMHD equations and 
high-resolution DNS on grids of $512^3$ and $1024^3$ points (see 
Table \ref{table:runs}). In these simulations, a power law close 
to $k^{-5/3}$ is observed before the kinetic energy spectrum drops to 
a steeper slope. As an example, Fig. \ref{fig:satura2} shows the kinetic 
and magnetic energy spectra in a simulation from set 9 using the LAMHD 
model, with $R_M \approx 41$ ($P_M = 0.0125$). Slopes corresponding to 
$k^{-5/3}$, $k^{-3}$, and $k^{-5}$ are indicated as a reference in Fig. 
\ref{fig:satura2}. A $k^{-5}$ power law in the magnetic energy spectrum 
(following a $k^{-3}$ range) was observed in experiments of dynamo action 
with constrained helical flows at low $R_M$ \cite{Muller04}; in addition, 
a $k^{-3}$ power law for the kinetic energy spectrum is consistent with 
the observed magnetic energy spectrum \cite{Leorat81}. Note that these 
power laws are only discussed here in order to be able to compare with 
the experimental data, but higher Reynolds numbers and thus more 
resolution will be needed in order to ascertain the spectral dependency 
of the flow and the magnetic field in the different inertial ranges of 
low $P_M$ simulations.

\begin{figure}
\includegraphics[width=9cm]{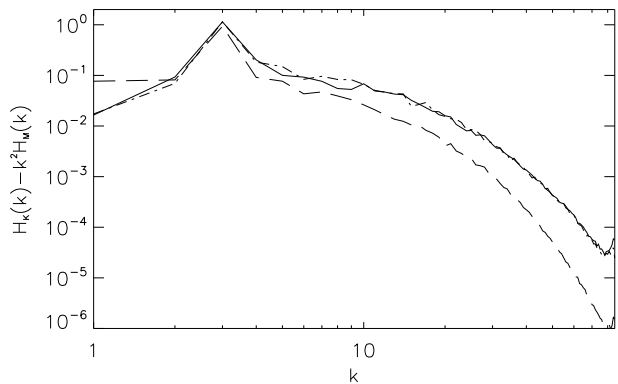}
\caption{Spectrum of $H_V(k)-k^2H_M(k)$, proportional to (minus) the 
    non-linear $\alpha$-effect, in the simulation in set 6 with 
    $R_M \approx 41$ ($P_M = 0.05$). The labels are as in Figs. 
    \ref{fig:spectrum} and \ref{fig:helspec}: the solid line is for 
    $t=11$, the dashed line for $t=29$, and the long dashed line for 
    $t=120$. Note the drop of the spectrum at late times at scales 
    smaller than $k_F$.}
\label{fig:alpha}
\end{figure}

Figure \ref{fig:helspec} shows the spectrum of relative magnetic helicity 
$k^{-1} H_M(k)/E_M(k)$ at different times for the same run as in Figs. 
\ref{fig:spectrum} and \ref{fig:satura1} (run with $R_M \approx 41$ in 
set 6). At all times, scales larger than the forcing scale 
have negative magnetic helicity, while scales of the order of, or smaller 
than the forcing scale have positive magnetic helicity. This is consistent 
with an inverse cascade of negative magnetic helicity at wavenumbers 
smaller than $k_F$, and with a direct transfer of positive magnetic 
helicity at wavenumbers larger than $k_F$, as analyzed in 
\cite{Alexakis06} using transfer functions. The relative helicity in 
the $k=1$ shell grows with time until reaching saturation. Note that 
at late times, $H_M(k=1)/E_M(k=1) \approx -1$, indicating that the 
large scale magnetic field is nearly force-free.

\begin{figure*}
\includegraphics[width=12cm]{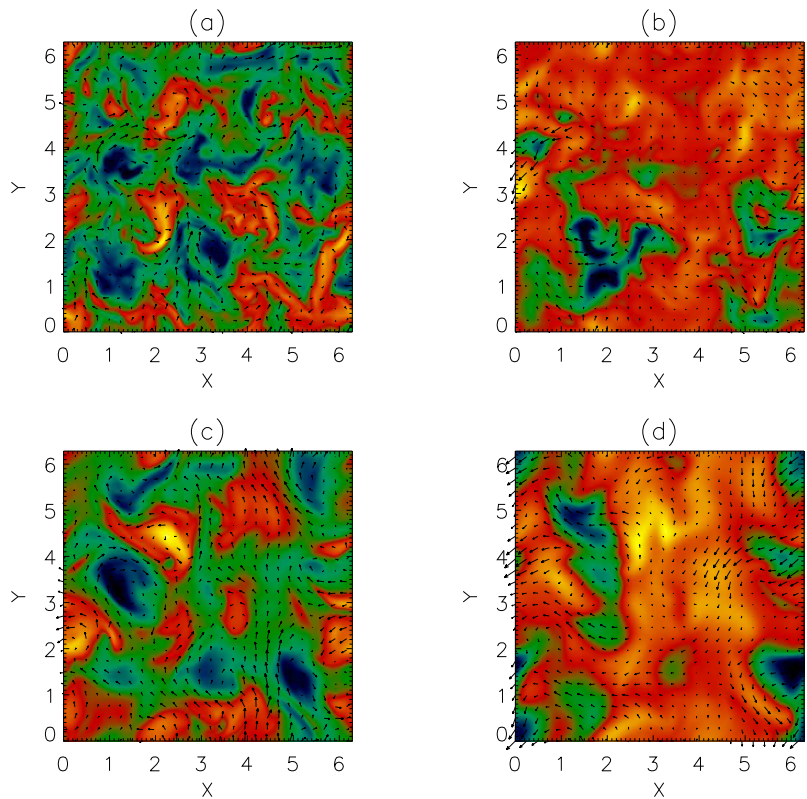}
\caption{(Color online) Plots of the velocity and magnetic fields in 
    a cut at $z=0$ for the simulation in set 6 with $R_M \approx 41$ and 
    $R_V\approx 820$ ($P_M = 0.05$): (a) $v_z$ component in color and 
    $v_x$, $v_y$ indicated by arrows at early time, (b) same as in (a) 
    for the magnetic field at early time, (c) same as in (a) at late 
    time, and (d) same as in (b) at late time.}
\label{fig:cut}
\end{figure*}

Figure \ref{fig:alpha} also shows the spectrum of $H_V(k)-k^2H_M(k)$, 
proportional to (minus) the non-linear $\alpha$-effect \cite{Pouquet76}. 
Three times are shown for the same run than in Figs. \ref{fig:spectrum}, 
\ref{fig:satura1}, and \ref{fig:helspec} (set 6, $R_M \approx 41$). At 
early times ($t = 11$ and $t=29$) the spectrum of $H_V(k)-k^2H_M(k)$ is 
close to the spectrum of the kinetic helicity. However, as the large 
scale magnetic field grows ($t=120$ is shown in the figure) the 
current helicity $k^2H_M(k)$ quenches kinetic helicity fluctuations 
and the total spectrum drops at scales smaller than $k_F$.

\begin{figure}
\includegraphics[width=9cm]{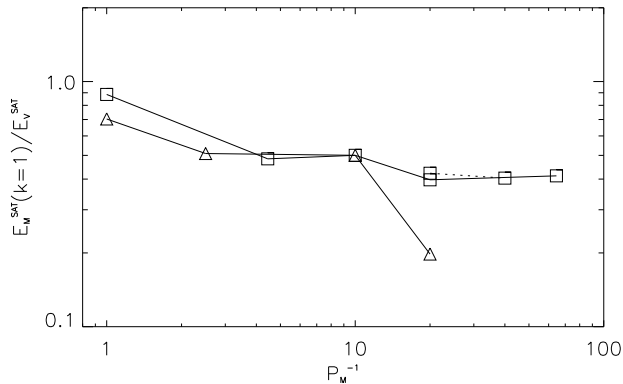}
\caption{Saturation value of the magnetic energy in the $k=1$ shell, 
     normalized by the total kinetic energy. The triangles correspond 
     to simulations at constant $R_V$, while the squares correspond to 
     simulations at constant $R_M$ (squares connected with solid lines 
     are from DNS, while squares connected with dotted lines are from 
     LAMHD simulations).}
\label{fig:largesc}
\end{figure}

As a result, at late times the magnetic energy is mostly in the modes 
with wavenumber $k=1$, which corresponds to the largest available scale 
in the system. In addition, the large scale magnetic field is force-free 
(maximum relative helicity with $H_M(k)=E_M(k)$ at $k=1$. Figure 
\ref{fig:cut} shows slices of the velocity and magnetic fields 
at early and late times. The growth of a large scale magnetic field 
and the quenching of turbulent velocity fluctuations in the saturated 
regime can be easily identified.

The situation resembles other inverse cascade situations that have 
been studied numerically, in which the fundamental $k=1$ mode dominates 
the dynamics at long times and its growth is only limited by its own 
dissipation rate \cite{Hossain83,Brandenburg01,Mininni03,Brandenburg05}. 
In helical dynamo simulations at $P_M = 1$ this behavior has also been 
observed, although it was speculated that for $P_M<1$ the inverse cascade 
of magnetic helicity and the generation of large scale fields should be 
quenched \cite{Brandenburg01}. In fact, the generation of magnetic energy 
at scales larger than the forcing scale is not quenched as $P_M$ is 
decreased. This is illustrated in Fig. \ref{fig:largesc}, which shows 
the ratio of the magnetic energy in the $k=1$ shell to the total 
kinetic energy in the saturated state as a function of $P_M$. 
Curves both at constant $R_V$ and constant $R_M$ are given. For constant 
$R_M$ and small $P_M$ the magnetic energy in the large scales seems 
to be independent of $P_M$ and $R_V$. The overall shape of the curves 
is similar to the curves in Fig. \ref{fig:scene}, indicating that at late 
times the evolution of the total magnetic energy is dominated by the 
magnetic field in the large scales.

\section{Conclusion}
We have shown in this paper that the phenomenon of inverse cascade of 
magnetic helicity, and the ensuing growth of large-scale magnetic 
energy together with a force-free magnetic field at large times, is 
present at low magnetic Prandtl number, down to $P_M=0.005$ in kinematic 
regime studies and down to $P_M=0.01$ in simulations up to the nonlinear 
saturation. The quenching of the velocity in the small scales, already 
observed in laboratory experiments, is also present. The augmentation 
of the critical magnetic Reynolds number as $R_V$ increases is less 
than in the non-helical case \cite{Schekochihin04a,Ponty05,Schekochihin05}, 
and even smaller than what was found for helical flows when the large-scale 
dynamo is not permitted, as e.g. for the Roberts flow at $k \approx 1$ 
\cite{Mininni05d}. The reason for this difference is that in the present 
study we allowed for enough scale separation between the forcing scale 
and the largest scale for helical large-scale dynamo action to develop. 
The results are in agreement with studies using mean-field theory and 
shell models to study both large- and small-scale dynamo action 
\cite{Frick06}. Large-scale helical dynamo action in the $P_M<1$ regime 
requires much smaller magnetic Reynolds numbers to work than small-scale 
dynamos.

The challenge remains, numerically, to be able to reach values of the 
magnetic Prandtl number comparable to those found in geophysics and 
astrophysics and in the laboratory, i.e. $P_M\approx 10^{-5}$. However, 
it is unlikely that the dynamo instability found here down to 
$P_M=0.005$ would disappear as $P_M$ is lowered further. An open 
question, of importance from the experimental point of view when 
dealing with turbulent liquid metals, is whether the critical magnetic 
Reynolds number $R_M^c$ will stabilize, for a given flow, at a value 
intermediate between what it is at $P_M=1$ and the peak of the curve 
(see Fig. \ref{fig:crmm}), or whether for large-scale helical dynamo 
action and extremely low values of $P_M$, it will go back down to the value 
it has at $P_M=1$. The data up to this day suggests the former, but on 
the other hand the study made in the context of two-point closures of 
turbulence \cite{Leorat81} suggests the latter. This also means that 
reliable models of turbulent flows in MHD must be developed in order 
that we can explore in a more systematic way the parameter space 
characteristic of the flows of interest, as for the geo-dynamo or 
the solar dynamo.

\begin{acknowledgments}
The author is grateful to D.C. Montgomery and A. Pouquet for valuable 
discussions and their careful reading of this manuscript. Computer 
time was provided by NCAR and by the National Science Foundation 
Terascale Computing System at the Pittsburgh Supercomputing Center. 
NSF-CMG grant 0327533 provided partial support for this work.
\end{acknowledgments}

\bibliography{ms}

\end{document}